\documentclass[12pt,english]{article}
\usepackage[latin9]{inputenc}
\usepackage{geometry}
\usepackage{verbatim}
\usepackage{amsmath}
\usepackage{amssymb}
\usepackage{esint}

\makeatletter
\usepackage{babel}

\usepackage{babel}

\makeatother

\usepackage{babel}
\begin{document}
\begin{center}
\textbf{\Large Higher Spin Cosmology }{\Large {} } 
\par\end{center}

\bigskip{}

\begin{center}
{\large Chethan KRISHNAN$^{a}$}%
\footnote{\texttt{\large chethan@cts.iisc.ernet.in}%
}{\large , Avinash RAJU$^{a}$}%
\footnote{\texttt{\large avinash@cts.iisc.ernet.in}%
}{\large , Shubho ROY$^{a}$}%
\footnote{{\large {}}\texttt{\large sroy@het.brown.edu}%
}{\large {} }\\
 {\large {} \vspace{0.2in}
 and Somyadip THAKUR$^{a}$}%
\footnote{\texttt{\large somyadip@cts.iisc.ernet.in}%
}{\large {} %Avinash RAJU$^a$\footnote{{\texttt{avinashraju777@gmail.com}}} \\
%\vspace{0.1in}
%and Somyadip THAKUR$^a$\footnote{{\texttt{smydp3thkr@gmail.com}}} 
} \vspace{0.1in}

\par\end{center}

\global\long\def\thefootnote{\arabic{footnote}}

%\vspace{0.2cm}

\begin{center}
$^{a}$ {Center for High Energy Physics\\
 Indian Institute of Science, Bangalore, India}\\

\par\end{center}

\noindent \begin{center}
\textbf{Abstract} 
\par\end{center}

We construct cosmological solutions of higher spin gravity in 2+1
dimensional de Sitter space. We show that a consistent thermodynamics
can be obtained for their horizons by demanding appropriate holonomy
conditions. This is equivalent to demanding the integrability of the
Euclidean boundary CFT partition function, and reduces to Gibbons-Hawking
thermodynamics in the spin-2 case. By using a prescription of Maldacena,
we relate the thermodynamics of these solutions to those of higher
spin black holes in AdS$_{3}$.

\vspace{1.6cm}
 \vfill{}

\section{Introduction}

In $2+1$ dimensions, pure gravity has no (perturbative) dynamics
because curvature is completely rigid. But despite the lack of any
gravitational attraction, gravity in 2+1 dimensions is non-trivial
-- black holes solutions were discovered by Banados, Teitelboim and
Zanelli (BTZ) \cite{Banados:1992wn} as quotients of AdS$_{3}$ \cite{Banados:1992gq}.
This fact makes 2+1 D gravity an excellent theoretical laboratory
for testing a variety nonperturbative issues in quantum gravity, without
the added complications of curvature dynamics which play a huge role
in higher dimensions. However, %interest in gravity in $2+1$ dimensions 
effort in this direction did not begin in earnest until the work of
Witten \cite{Witten:1988hc} (see also \cite{Achucarro:1987vz}).
He demonstrated that $2+1$-d gravity %with or without a cosmological constant $\Lambda$ 
can be recast as a Chern-Simons gauge theory, with the gauge gauge
group $SL(2,R)\times SL(2,R)$ when the cosmological constant $\Lambda$
is $<0$, %or AdS$_{3}$
and the gauge group $SL(2,C)$ when $\Lambda$ is $>0$. %or dS$_{3}$). 

The negative cosmological constant case drew a lot of attention, partly
because that was the context in which the above mentioned BTZ black
holes were discovered, but also because of the earlier work of Brown
and Henneaux \cite{Brown:1986nw} who showed that the asymptotic symmetry
algebra of AdS$_{3}$ gravity is a Virasoro algebra. %i.e. the conformal group in 1+1d,  drew a lot of attention to AdS$_{3}$ case. 
In fact, this latter result is now widely recognized as a precursor
to the celebrated AdS-CFT duality \cite{Maldacena:1997re} where a
fully $quantum$ theory of gravity in AdS$_{d+1}$ is conjectured
to have an equivalent description in terms of a conformal field theory
supported on the boundary of AdS$_{d+1}$. However, there was no such
happy ending in the case of a positive cosmological constant, namely
dS$_{3}$. Although the counterpart to the BTZ black hole quotients
were constructed %and dubbed Kerr-dS$_{3}$
in \cite{Park:1998qk} (also see \cite{Balasubramanian:2001nb}),
the fact that de Sitter is a cosmological spacetime with a spacelike
boundary \cite{Spradlin:2001pw} has made the development of a consistent
dS/CFT proposal much more confusing. %and the asymptotic symmetry analysis yielded a parallel non-unitary
%version of the Virasoro algebra \cite{Spradlin:2001pw}, these efforts never culminated into any concrete or consistent dS/CFT kind of full quantum equivalence - the quantum dynamics of gravity in de Sitter space cannot be described by a CFT. 
Various interesting attempts were made in \cite{Witten:2001kn,Strominger:2001gp,Maldacena:2002vr},
but there seems to be a fundamental difficulty in realizing de Sitter
space in $any$ kind of unitary quantum set up as a stable vacuum
\cite{Polyakov:2007mm,Banks:2005bm,Fischler:2001yj,Dyson:2002nt,Dyson:2002pf,Krishnan:2006bq}.

On an entirely different theme, theories of interacting gauge fields
with an infinite tower of higher spins ($s\geq2$) have been studied
as a toy version of a full string theory%
\footnote{Since the latter has infinite dimensional gauge invariance, see the
work by Sundborg \cite{Sundborg:2000wp}.%
} by Fradkin and Vasiliev \cite{Fradkin:1986ka,Fradkin:1987ks,Vasiliev:1990cm,Vasiliev:1990en,Vasiliev:1992av,Vasiliev:1999ba},
building on the early work of Fronsdal \cite{Fronsdal:1978rb}. Higher
spin theories in three dimensions, as demonstrated in \cite{Blencowe:1988gj}
are considerably simpler than theories in higher dimensions due to
absence of any local propagating degrees of freedom. %i.e. they are topological. 
In addition, it is possible to truncate the infinite tower of higher
spin fields to spin, $s\leq N$. The complicated nonlinear interactions
of the higher spin fields can be reformulated in terms of an $SL(N,R)\times SL(N,R)$
Chern-Simons gauge theory (for AdS$_{3}$ case) or an $SL(N,C)$ Chern-Simons
(for dS$_{3}$). Therefore 2+1 dimensional higher spin theories are
a generalization of Chern- Simons gravity -- one gets back to the
spin-2 pure gravity theory when one sets $N=2$.

The aim of this paper is to construct cosmological solutions in higher
spin dS$_{3}$ gravity. We work specifically with the case where the
rank of the gauge group, $N=3$. The solutions we construct are the
higher spin generalizations of dS$_{3}$ quotients such as Kerr-dS$_{3}$
and quotient cosmology \cite{Park:1998qk,Balasubramanian:2001nb,Krishnan:2013cra}
and should be thought of as the de Sitter counterparts of the spin-3
charged AdS$_{3}$ black hole solutions of \cite{Gutperle:2011kf,Ammon:2011nk}.
It has been shown recently by two of us that big-bang type singularities
contained in quotient cosmologies in the purely $SL(2)$ sector of
this higher spin theory can be removed by performing a spin-3 gauge
transformation \cite{Krishnan:2013cra}. But the problem of constructing
spin-3 charged cosmologies was left open. In this paper, we fill this
gap and discuss the thermodynamics of their cosmological horizons.

The plan of the paper is the following. In Sec. \ref{sec:Basic SL(3,C) CS},
we recap the formulation of spin-3 field coupled to gravity in 2+1-dimensions
as a Chern-Simons gauge theory with a non-compact gauge group, $SL(3,C)$.
We fix all notations and conventions for the map between the second
order variables (metric and spin-3 field) and the gauge connection
here. We also review and discuss the variational principle for asymptotically
de Sitter like connections in the gauge theory formulation. In Sec.
\ref{sec:hs dS cosmologies}, we first review pure gravity i.e. $SL(2,C)$
sector solutions, % in asymptoticallyde Sitter spaces. T
namely the Kerr de Sitter universe and the quotient cosmology. We
do this both in the metric and gauge theory set-up. Then we construct
higher spin extensions of these geometries by modifying their gauge
connection and adding spin-3 charges in a manner consistent with the
triviality of gauge connection holonomies along contractible cycles.
These solutions are shown to contain cosmological horizons and, in
the case of quotient cosmology, higher spin big bang/ big crunch like
causal singularities. In the final section, Sec. \ref{sec: dS TD},
these holonomy conditions are shown to be necessary for the consistency
%\textbf{(what does consistent thermodynamics mean? What lets us know that we scored?)} 
of thermodynamics associated with cosmological horizons. These consistency
conditions turn out to be identical to demanding integrability of
a ``boundary CFT partition function''. Using a prescription of Maldacena
\cite{Maldacena:2002vr}, we relate thermodynamics of our solutions
to those of higher spin AdS$_{3}$ black holes. % through a ``wick-rotation'' of gauge connectionfrom dS to AdS. 
Our formulation gives the same results as the Gibbons-Hawking results
when we restrict to spin-2 and work in the metric language.\\

%\textbf{Cite Stephane and Gaston.}

\section{$SL(3,C)$ Chern-Simons formulation of Higher Spin dS$_{3}$-gravity\label{sec:Basic SL(3,C) CS}}

Here we quickly review the basics of the $SL(3,C)$ CS gauge theory
generalizing Witten's construction \cite{Witten:1988hc} as presented
in detail in \cite{Ouyang:2011fs,Krishnan:2013cra} (see also \cite{Lal:2012py}
for discussion on higher spins in dS$_{3}$). One simply defines the
higher spin (up to spin 3) theory i.e. an interacting theory of gravity
and a spin 3-field, by the action \cite{Blencowe:1988gj} 
\begin{equation}
I_{CS}[A]=\frac{k}{4\pi d_{R}}\int_{M}\mbox{Tr}\left(AdA+\frac{2}{3}A^{3}\right)-\frac{k}{4\pi d_{R}}\int_{M}\mbox{Tr}\left(\bar{A}d\bar{A}+\frac{2}{3}\bar{A}^{3}\right).\label{eq: CS action for dS_3}
\end{equation}
The constant $d_{R}=-2\mbox{Tr}\left(T_{0}T_{0}\right)$ is a characteristic
of the representation size.  Here the gauge field, $A$ is a complex $SL(3)$ matrix-valued one
form. In a basis of $SL(3)$ matrices, $\{T_{a},T_{ab};a,b=0,1,2\}$
as listed in \cite{Campoleoni:2010zq}, we can expand the gauge field
as 
\begin{eqnarray*}
A & = & \left(\omega_{\mu}^{a}+\frac{i}{l}e_{\mu}^{a}\right)T_{a}dx^{\mu}+\left(\omega_{\mu}^{ab}+\frac{i}{l}e_{\mu}^{ab}\right)T_{ab}dx^{\mu}\\
 & = & \left(\omega_{\mu}+\frac{i}{l}e_{\mu}\right)dx^{\mu},\;\omega_{\mu}=\omega_{\mu}^{a}T_{a}+\omega_{\mu}^{ab}T_{ab},e_{\mu}=\omega_{\mu}^{a}T_{a}+\omega_{\mu}^{ab}T_{ab}.
\end{eqnarray*}
 Then, the more familiar metric and spin-3 field can be extracted
from the (imaginary parts of the basis coefficients) of the gauge
field \cite{Campoleoni:2010zq}: 
\begin{equation}
g_{\mu\nu}=\frac{1}{2!}\mbox{Tr}\left(e_{\mu}e_{\nu}\right),\phi_{\mu\nu\lambda}=\frac{1}{3!}\mbox{Tr}\left(e_{(\mu}e_{\nu}e_{\lambda)}\right),\label{eq: metric and spin-3 field definition}
\end{equation}
 while the three-dimensional Newton's constant (in units of the dS
radius, $l$) is given by the Chern-Simons level number, 
\begin{equation}
\frac{G_{3}}{l}=\frac{1}{4 i k}.\label{eq: Relation between CS level and Newton's constant}
\end{equation}
 %The constant $d_{R}=-2\mbox{Tr}\left(T_{0}T_{0}\right)$ is a characteristic of the representation size. 
We work in the prevalent general relativity
convention where, $8G_{3}=1.$ Since the gauge group $SL(3,C)$ is
non-compact the Chern-Simons level number is $not$ quantized.

Now lets consider the variation of the action (\ref{eq: CS action for dS_3}).
Generically a variation has a bulk (volume) piece, proportional to
the equation of motion and boundary pieces supported on temporal and
spatial boundaries, 
\begin{eqnarray}
\delta I=\int d^{3}x\:(E.O.M)+\int d^{2}x\,\pi_{\mu}\delta A^{\mu}|_{t_{i}}^{t_{f}}+\int dtdx^{j}\;\left.\pi_{\mu}^{j}\delta A^{\mu}\right|_{x_{i,min}}^{x_{i,max}}
\end{eqnarray}
 To have a good variational principle one has to ensure that these
boundary pieces vanish (on-shell) by prescribing initial and final
conditions and spatial boundary conditions. If the prescribed conditions
do not lead to vanishing contribution for the boundary pieces of the
variation, then one has to add supplementary boundary terms to the
action to cancel these. One crucial point to be noted here in contrast
with the AdS case is that the action (\ref{eq: CS action for dS_3})
already defines a good variational principle without any supplementary
boundary terms. This is because asymptotically deSitter spaces have
$closed$ spatial sections and the only boundary contributions are
from future infinity ($t_{f}\rightarrow\infty$) and at some time
coordinate in the past ($t_{i}$ $={\rm const}$). As the variational
principle is usually defined with vanishing variations at the initial
and final times, 
\begin{eqnarray}
\delta A|_{t_{i},t_{f}}=0,
\end{eqnarray}
 these boundary pieces vanish. However we shall $not$ demand the
future data to be fixed (i.e. $\delta A|_{t_{f}\rightarrow\infty}\neq0$)
and look to set up a variational principle by demanding instead the
conjugate momentum vanishes 
\begin{eqnarray}
\pi_{\mu}|_{t_{f}\rightarrow\infty}\rightarrow0.
\end{eqnarray}
 Such a variational principle will be made to appear natural in Sec.
\ref{sec: dS TD} where the close parallel between de Sitter and Anti
de Sitter cases is brought out. This will often restrict us to a subclass
of solutions which are specified by their future fall-off behaviors
(which close under gauge transformations), 
\begin{eqnarray}
\lim_{t\rightarrow\infty}A_{\mu}\sim t^{\alpha_{\mu}}
\end{eqnarray}
 for some real $bounded$ exponent $\alpha_{\mu}$. This is the analogue
of non-normalizable fall-offs in AdS. These fall-off behaviors are
fixed by conducting the asymptotic (future/past) symmetry analysis
in a manner closely parallel to the AdS$_{3}$ counterpart \cite{Campoleoni:2010zq,Henneaux:2010xg}
as was done in \cite{Ouyang:2011fs}. By demanding that the asymptotic
symmetries of this larger theory still contain the Virasoro algebras
already present in the $SL(2,C)$ case, it was found that the suitable
fall-offs behaviors at future infinity for the $SL(3,C)$ gauge connections
are 
\begin{equation}
A_{\bar{w}}=0,\qquad A_{\rho}=b^{-1}\partial_{\rho}b,\qquad A-A_{dS_{3}}\stackrel{\tau\rightarrow\infty}{\longrightarrow}\mathcal{O}(1).\label{eq: asymptotic future dS_3}
\end{equation}
 Here $b$ is a gauge transformation $\in SL(3,C)$. However as we
shall see in the next section, in order to construct gauge field configurations
with non-vanishing higher spin charges, one has to violate the asymptotic
fall-offs (\ref{eq: asymptotic future dS_3}) and hence one has to
supplement the action (\ref{eq: CS action for dS_3}) with boundary
terms. Again this is parallel to the situation for higher spin AdS
black hole solutions \cite{Gutperle:2011kf} for which the boundary
counter terms were worked out in \cite{Banados:2012ue,deBoer:2013gz}.

\section{Higher Spin de Sitter Cosmologies\label{sec:hs dS cosmologies}}

We are interested in constructing solutions of the $SL(3,C)$ gauge
theory describing spacetimes of positive cosmological constant which
have non-zero spin-3 charges in addition to the spin-2 charges i.e.
energy and angular momentum. Since these are higher spin extensions
of the pure gravity solutions or $SL(2)$ sector, let us first review
the solutions of the $SL(2)$ sector obtained by taking quotients
of pure three-dimensional de Sitter space \cite{Balasubramanian:2001nb}.

\subsection{Kerr-dS$_{3}$ universe}

The first class of $SL(2)$ quotients of pure de Sitter space is the
so called Kerr - deSitter universe ($KdS_{3}$). These are very similar
to de Sitter space itself, in the sense that these solutions have
two regions bounded by cosmological horizons, and have future and
past infinite regions outside the cosmological horizons. However the
topology of the past and future infinities of $KdS_{3}$ is that of
a cylinder, $S^{1}\times R$, in contrast to the de Sitter space,
for which they have topology of a sphere, $S^{2}$. %The Penrose diagram of Kerr de Sitter universe is provided in FIG. XYZ.

$\,$

In static Schwarzschild-like coordinates, the $KdS{}_{3}$ metric
\cite{Park:1998qk,Balasubramanian:2001nb} reads like, 
\begin{equation}
ds^{2}=-N^{2}(r)dt^{2}+N^{-2}(r)dr^{2}+r^{2}\left(d\phi+N_{\phi}dt\right)^{2},N^{2}(r)=M-\frac{r^{2}}{l^{2}}+\frac{J^{2}}{4r^{2}},N_{\phi}=-\frac{J}{2r^{2}}.\label{eq: Kerr-dS_3 in Sch}
\end{equation}
 Introducing, the outer and inner radii 
\begin{equation}
r_{\pm}^{2}=Ml^{2}\left(\sqrt{1+\left(J/M\, l\right)^{2}}\pm1\right)/2,\label{eq: r_plus and r_minus}
\end{equation}
 one can rewrite Eq.(\ref{eq: Kerr-dS_3 in Sch}) as 
\begin{equation}
ds^{2}=-\frac{\left(r^{2}+r_{-}^{2}\right)\left(r_{+}^{2}-r^{2}\right)}{r^{2}l^{2}}dt^{2}+\frac{r^{2}l^{2}}{\left(r^{2}+r_{-}^{2}\right)\left(r_{+}^{2}-r^{2}\right)}dr^{2}+r^{2}\left(d\phi+\frac{r_{+}r_{-}}{r^{2}}\frac{dt}{l}\right)^{2},r<r_{+}\label{eq:Kerr-dS_3 inside}
\end{equation}
 and we note that this geometry has a horizon at $r=r_{+}$. This
metric can be analytically continued across outside i.e. for $r>r_{+}$:
\begin{equation}
ds^{2}=-\frac{r^{2}l^{2}}{\left(r^{2}+r_{-}^{2}\right)\left(r^{2}-r_{+}^{2}\right)}dr^{2}+\frac{\left(r^{2}+r_{-}^{2}\right)\left(r^{2}-r_{+}^{2}\right)}{r^{2}l^{2}}dt^{2}+r^{2}\left(d\phi+\frac{r_{+}r_{-}}{r^{2}}\frac{dt}{l}\right)^{2}.\label{eq:Kerr-dS_3 outside}
\end{equation}
 In this region $r$ is timelike while $t$ is spacelike.\\

To make contact with the gauge theory we write down the $SL(2,C)$
connections for the two regions. Introducing, $\mathcal{N}^{2}(r)\equiv\frac{\left(r^{2}+r_{-}^{2}\right)\left(r^{2}-r_{+}^{2}\right)}{r^{2}l^{2}}=-N^{2}(r)$,
the gauge field expressions are,

$ $ 
\begin{eqnarray}
A^{0} & = & N(r)\left(d\phi+i\frac{dt}{l}\right),\qquad A^{1}=\frac{lN_{\phi}-i}{N(r)}\frac{dr}{l}\qquad,A^{2}=\left(rN_{\phi}+i\frac{r}{l}\right)\left(d\phi+i\frac{dt}{l}\right);\qquad r<r_{+},\nonumber \\
A^{0} & = & -\frac{lN_{\phi}-i}{\mathcal{N}(r)}\frac{dr}{l},\qquad A^{1}=\mathcal{N}(r)\left(d\phi+i\frac{dt}{l}\right),\qquad A^{2}=\left(rN_{\phi}+i\frac{r}{l}\right)\left(d\phi+i\frac{dt}{l}\right);\qquad r>r_{+}.\label{eq: SL(2) gauge connection for KdS_3 in Sch.}
\end{eqnarray}
 \\

In the exterior region, $r>r_{+}^{2}$, on can transform to Fefferman-Graham
like coordinates ($\rho,w,\bar{w}$) defined by 
\begin{equation}
\rho=\ln\left(\frac{\sqrt{r^{2}-r_{+}^{2}}+\sqrt{r^{2}+r_{-}^{2}}}{2l}\right),w=\phi+it/l,\bar{w}=\phi-it/l\label{eq: FG radius in terms of Sch. radius}
\end{equation}
 and obtain the form of the metric, 
\begin{equation}
ds^{2}=-l^{2}d\rho^{2}+\frac{1}{2}\left(Ldw^{2}+\bar{L}d\bar{w}^{2}\right)+\left(l^{2}e^{2\rho}+\frac{L\bar{L}}{4}e^{-2\rho}\right)dwd\bar{w},\label{eq: Kerr-dS_3 in FG}
\end{equation}
 where, the zero modes, $L,\bar{L}$ are defined by, 
\begin{equation}
L+\bar{L}=Ml,\qquad L-\bar{L}=iJ.\label{eq: Zero Modes}
\end{equation}
 Note that $\rho$ here is a time coordinate, eg. \cite{deBuyl:2013ega}.

This coordinate system is better suited than the Schwarzschild one
for conducting the asymptotic symmetry analysis of dS$_{3}$ and its
identification with Euclidean Virasoro algebra and its charges \cite{Balasubramanian:2001nb}.
As we have worked out in our previous paper \cite{Krishnan:2013cra},
the corresponding $SL(2,C)$ gauge field is, 
\begin{equation}
A=i\, T_{0}\, d\rho+\left[\left(e^{\rho}-\frac{L}{2l}e^{-\rho}\right)\, T_{1}+i\,\left(e^{\rho}+\frac{L}{2l}e^{-\rho}\right)T_{2}\right]dw\label{eq: SL(2,C) connection for general metric}
\end{equation}
 One can obtain the above Kerr-dS$_{3}$ connection from a primitive
connection, $a$ given by 
\begin{equation}
a=\left[\left(1-\frac{L}{2l}\right)T_{1}+i\left(1+\frac{L}{2l}\right)T_{2}\right]dw\label{eq: Kerr-dS3 curly}
\end{equation}
 free of any $\rho$ dependence, by performing a single valued gauge
transformation on $a$: 
\begin{eqnarray}
A=\mathcal{B}^{-1}a\:\mathcal{B}+\mathcal{B}^{-1}d\mathcal{B},
\end{eqnarray}
 for 
\begin{equation}
\mathcal{B}=\exp\left(i\rho T_{0}\right)=\exp(\rho L_{0})\label{eq: Radial gauge transformation}
\end{equation}
 (because $\mathcal{B}$ being a sole function of $\rho$ is single
valued in the $\phi$ direction).

\subsection{Quotient Cosmology}

One can also construct de Sitter quotients containing (spinning) big
bang/big crunch singularities \cite{Balasubramanian:2001nb} (also
reviewed in \cite{Castro:2012gc} ). These quotients are locally given
by the same exterior Kerr- de Sitter metric (\ref{eq:Kerr-dS_3 outside}).
But since $t$ and $r$ switch their roles and become spacelike and
timelike respectively, we are better off switching their roles in
the metric itself, 
\begin{equation}
ds^{2}=-\frac{t^{2}l^{2}}{\left(t^{2}+r_{-}^{2}\right)\left(t^{2}-r_{+}^{2}\right)}dt^{2}+\frac{\left(t^{2}+r_{-}^{2}\right)\left(t^{2}-r_{+}^{2}\right)}{t^{2}l^{2}}dr^{2}+t^{2}\left(d\phi+\frac{r_{+}r_{-}}{t^{2}}dr\right)^{2}.\label{eq: Quotient Cosmology metric}
\end{equation}
 The quotient cosmology arises when we compactify $r$ into a circle.
%with periodicity $2\pi\frac{l^{2}r_{+}}{r_{+}^{2}+r_{-}^{2}}$. 
With,
$r$ and $\phi$ both being periodic the future and past infinity
of this quotient cosmology have the topology of a torus, $S^{1}\times S^{1}$
as opposed to $R\times S^{1}$ for the case of the Kerr de Sitter
universe. Also with a periodic $r$, this metric cannot be extended
to $-r_{+}<t<r_{+}$ where $g_{rr}<0$ and one has closed timelike
curves. Removing this region then leaves us with a big bang (big crunch)
like solution for $t>r_{+}$ ($t<r_{+}$) with the $r$-$\phi$ torus
degenerating to a circle \cite{Balasubramanian:2001nb}. This is an example of a causal structure singularity \cite{Banados:1992gq}, and these are the analogues of higher dimensional curvature singularities in 2+1 dimensions. %The Penrose diagram for de Sitter quotient cosmology is provided in FIGURE 2.
These singularities were shown to be removable via a higher spin gauge
transformation when we embed this metric into a spin-3 $SL(3)$ theory
in \cite{Krishnan:2013cra}.

$\,$

Since the quotient cosmology is metrically identical to the exterior
regions of the Kerr de Sitter universe, the Fefferman-Graham gauge
metric expression (\ref{eq: Kerr-dS_3 in FG}) and the gauge connection
expressions (\ref{eq: SL(2,C) connection for general metric},\ref{eq: Kerr-dS3 curly},\ref{eq: Radial gauge transformation})
also carry over with the coordinate changes, 
\begin{equation}
\rho=\ln\left(\frac{\sqrt{t^{2}-r_{+}^{2}}+\sqrt{t^{2}+r_{-}^{2}}}{2l}\right),w=\phi+ir/l,\bar{w}=\phi-ir/l.\label{eq: FG time in terms of Schwarzschild time}
\end{equation}

\subsection{The Higher Spin cosmological gauge fields}

\noindent %Following the approach of \cite{Campoleoni:2010zq,Ouyang:2011fs} we now turn on the spin-3 sector of gauge field demanding asymptotic (future) de Sitter fall conditions on the connection. 
In the $SL(3)$ theory, the general primitive connection that satisfies
asymptotic (future) de Sitter fall off conditions is %gauge connection has the form 
\begin{equation}
a'=\left[\left(1-\frac{L}{2l}\right)T_{1}+i\left(1+\frac{L}{2l}\right)T_{2}+\frac{W}{8l}W_{-2}\right]dw\label{eq:Asymptotic de Sitter form}
\end{equation}
 $L$ and $W$ can be functions of $z$, but we will consider the
constant case in analogy with \cite{Gutperle:2011kf}. Explicit forms
for the generators can be found in \cite{Krishnan:2013cra}.

%\textbf{Mention where we define $W$'s.}
$\,$ We can transform from the primitive connection $a'$, to $A'$,
the fully $\rho$-dependent form by applying the transformation (\ref{eq: Radial gauge transformation})

\noindent 
\begin{equation}
A'=iT_{0}d\rho+\left[\left(e^{\rho}-\frac{L}{2l}e^{-\rho}\right)\, T_{1}+i\,\left(e^{\rho}+\frac{L}{2l}e^{-\rho}\right)T_{2}+\frac{W}{8l}e^{-2\rho}W_{-2}\right]dw,\label{eq: dS fall-off with spin-3}
\end{equation}

\noindent which we call the Fefferman-Graham gauge because it manifests
the proper $\rho\rightarrow\infty$ fall-offs behaviors Eq. (\ref{eq: asymptotic future dS_3})
as derived in \cite{Campoleoni:2010zq,Ouyang:2011fs}. \\

\noindent As the trace of $W_{-2}$ with any $SL(3)$ generator is
zero, we find that metric obtained from $A',\bar{A}'$ is same as
(\ref{eq: Kerr-dS_3 in FG}). But the spin-3 field now attains a non-zero
value. These non-vanishing components of spin-3 fields are given by

\begin{align}
\varphi_{www} & =-\frac{i}{8}l^{2}W,\nonumber \\
\varphi_{ww\bar{w}} & =-\frac{i}{24}l\bar{L}We^{-2\rho}+\frac{i}{24}l^{2}\overline{W},\nonumber \\
\varphi_{w\bar{w}\bar{w}} & =-\frac{i}{96}\bar{L}^{2}We^{-4\rho}+\frac{i}{24}l\bar{L}\overline{W}e^{-2\rho},\nonumber \\
\varphi_{\bar{w}\bar{w}\bar{w}} & =\frac{i}{32}\bar{L}^{2}\overline{W}e^{-4\rho}.\label{eq:Spin-3 field components}
\end{align}

In order to construct metrics (cosmologies) with $nonvanishing$ spin-3
charges (which will necessarily violate the asymptotically dS fall-offs),
we propose the following ansatz for the primitive connection corresponding
to a general spin-3 cosmology

\begin{equation}
a'=\left[\left(1-\frac{L}{2l}\right)T_{1}+i\left(1+\frac{L}{2l}\right)T_{2}+\frac{W}{8l}W_{-2}\right]dw+\mu\left[W_{2}+w_{0}W_{0}+w_{-2}W_{-2}+t\;(T_{1}-iT_{2})\right]d\bar{w},\label{eq: primitive spin-3 cosmology}
\end{equation}

\noindent where $\mu,w_{0},w_{-2},$ and $t$ are constants. The motivation
for this comes from the fact that under a suitable set of analytically
continuations of the charges and sign of the cosmological constant
(which will be elaborated in the following sections) de Sitter higher
spin cosmologies turn into the Euclidean sections of AdS higher spin
black hole solutions of \cite{Gutperle:2011kf} (much like in the
case of pure gravity or $SL(2)$ sector, Kerr-dS$_{3}$ solutions
continue on to Euclidean $BTZ$ black holes). \\

\noindent Now, the connection (\ref{eq: primitive spin-3 cosmology})
is an off-shell object and contains too many independent parameters.
Restricting on-shell, we find that the connection has to be of the
form

\begin{equation}
a'=\left[\left(1-\frac{L}{2l}\right)T_{1}+i\left(1+\frac{L}{2l}\right)T_{2}+\frac{W}{8l}W_{-2}\right]dw+\mu\left[W_{2}-\frac{L}{2l}W_{0}+\frac{L^{2}}{16l^{2}}W_{-2}+\frac{W}{l}(T_{1}-iT_{2})\right]d\bar{w}.\label{eq: On-shell primitive spin-3}
\end{equation}

\noindent Now although the connection is on-shell, it is still arbitrary
in the sense that one does $not$ know whether such solutions make
a regular or singular contribution to the Hartle-Hawking wave-function
(or equivalently, when continued to Euclidean AdS, the corresponding
Gibbons-Hawking partition function, $Z_{ECFT}$). Just as in the second-order
or metric formulation of gravity, this is guaranteed by demanding
the regularity of the Euclidean section of the metric, in case of
the first-order or connection formulation, it is fixed by demanding
$triviality$ of the gauge-connection $A$ along contractible circle(s).
The non-trivial topology of the connection is captured by the holonomy
matrix or the Wilson loop operator along any contractible circle,
$\mathcal{C}$

\begin{eqnarray}
\mathrm{Hol}_{\mathcal{C}}(A)\equiv\mathcal{B}^{-1}\exp\left[\oint_{\mbox{\ensuremath{\mathcal{C}}}}\; dx^{\mu}a_{\mu}\right]\;\mathcal{B}=e^{H_{\mathcal{C}}}.
\end{eqnarray}

\noindent The triviality of the connection is ensured when this holonomy matrix is identity. Equivalently, this
means that the matrix, $H_{\mathcal{C}}$ has eigenvalues $(0,-2\pi i,2\pi i)$.
In the case of Kerr-dS$_{3}$ and its spin-3 generalizations one has
a contractible thermo-angular circle, 
\begin{eqnarray}
\left(t,\phi\right)\sim\left(t+i\beta,\phi+i\beta\Omega\right).
\end{eqnarray}
 Equivalently, if one defines $\tau=\frac{\beta}{2\pi}\left(1-i\Omega l\right)$,
this thermo-angular circle can be reexpressed as $\left(w,\bar{w}\right)\sim\left(w+2\pi\tau/l,\bar{w}+2\pi\bar{\tau}/l\right).$\\
 The associated holonomy, 
\begin{eqnarray}
\mbox{Hol}(a) & = & \mathcal{B}^{-1}\exp\left[\int_{0}^{i\beta}\; dt\, a_{t}+\int_{0}^{i\beta\Omega}\: d\phi\: a_{\phi}\right]\;\mathcal{B}\nonumber \\
 & = & \mathcal{B}^{-1}\exp\left[\int_{0}^{i\beta}\; dt\, i\left(a_{w}-a_{\bar{w}}\right)/l+\int_{0}^{i\beta\Omega}\: d\phi\:\left(a_{w}+a_{\bar{w}}\right)\right]\;\mathcal{B}\nonumber \\
 & = & \mathcal{B}^{-1}\exp\left[-\left(2\pi\tau\: a_{w}-2\pi\bar{\tau}\: a_{\bar{w}}\right)/l\right]\;\mathcal{B}\nonumber \\
 & = & e^{W(a)}.\label{eq: Holonomy around the thermo-angular circle}
\end{eqnarray}
 This means that the matrix $W(a)$ should have eigenvalues $(0,-2\pi i,2\pi i)$%
\footnote{Of course, one could consider the eigenvalues to be integer multiples
of $\pm2\pi i$, in general, to get trivial holonomy. This ambiguity is directly
tied to the ambiguity in identifying the period of the thermal circle
with the inverse temperature, which in turn is ultimately tied to fixing the asymptotics of the geometry \cite{Krishnan:2010un}. %For holonomy eigenvalues which are chosen to be integer multiples of $2\pi$, i.e. lets say $2\pi K$, one has to identify the inverse temperature with $\tau'\equiv\tau/|K|$. %A  related, but distinct, subtlety is the relevance of the center of the gauge group in determining the holonomy. A careful discussion of this question can be found in \cite{Castro:2011fm}.%
The choice $(0,-2\pi i,2\pi i)$ can be found in section 5.4 of \cite{Gutperle:2011kf}. It is easy to see from (30) that once we fix a period $\tau$, scaling the holonomy eigenvalues by $N$ can only be accomplished by scaling the primitive connection $a$ in (27). But this results in a metric (and connection $A$) that will violate the standard Brown-Henneaux fall-offs (and their higher spin generalizations). This is again ultimately tied to the fact that the asymptotic fall-offs are defined after fixing the asymptotic coordinates; the Killing vectors are normalized at infinity. Eg: note that the norm of the Killing vector that turns null at the horizon is crucial for determining the surface gravity/temperature \cite{Krishnan:2010un}. %The Brown-Henneaux Virasoro is precisely the left over diffeomoprhisms of the boundary.
}. These two (complex) conditions entirely fix the charges $L,W$ in
terms of the potentials $\beta,\mu$.\\

\noindent For generic gauge connections it is nontrivial to compute
the holonomy matrix exactly. Since all we need are its eigenvalues,
we are perfectly fine to work with the matrix, $\exp(\tilde{w}(a)),\:\tilde{w}(a)\equiv-\left(2\pi\tau\, a_{w}-2\pi\bar{\tau}\: a_{\bar{w}}\right)$,
instead, since it is related to the Holonomy matrix, $\exp(W(a))$
by a single-valued gauge transformation (similarity transformation),
$\mathcal{B}$ and hence has the same eigenvalue spectrum. Demanding
the eigenvalues of the $\tilde{w}_{z}$ be $(0,-2\pi i,2\pi i)$ implies

\begin{equation}
\det\left(\tilde{w}(a)\right)=0,\quad\text{Tr}\left[\tilde{w}(a)^{2}\right]=-8\pi^{2},\quad\text{Tr}\left[\tilde{w}(a)\right]=0.
\end{equation}
 which translate to the following relations determining the charges
$L,W$ in terms of the potentials $\tau,\mu$,

\noindent 
\begin{eqnarray}
27l^{2}\tau^{3}W-36l\tau^{2}\alpha L^{2}-54l\tau\alpha^{2}LW-54l\alpha^{3}W^{2}-8\alpha^{3}L^{3} & = & 0\nonumber \\
1-\frac{2\tau^{2}L}{l^{3}}-\frac{6\tau\alpha W}{l^{3}}+\frac{4}{3}\frac{\alpha^{2}L^{2}}{\tau^{2}l^{2}}\qquad\qquad\qquad & = & 0,\label{eq:Holonomy constraints}
\end{eqnarray}
 where $\alpha=\mu\bar{\tau}$. \\

\noindent Since we are already familiar with the exact solution for
purely spin-2 charges i.e. mass and angular momentum, we can now obtain
a solution to the charges in the presence of spin-3 potentials in
a perturbation series in the spin-3 chemical potential, $\mu$:

\begin{equation}
W=\sum_{i=1}^{\infty}a_{i}\mu^{i}\qquad L=\frac{l}{2\tau^{2}}+\sum_{j=1}^{\infty}b_{j}\mu^{j}
\end{equation}

\noindent Substituting this in Eq.(\ref{eq:Holonomy constraints})
and solving both equations to quadratic order $\mu$, we get following
perturbative solution for $L$ and $W$ 
\begin{equation}
\frac{L}{l}=\frac{l^{2}}{2\tau^{2}}-\frac{5\alpha^{2}l^{4}}{6\tau^{6}}+\cdots,\qquad\frac{W}{l}=\frac{\alpha l^{4}}{3\tau^{5}}-\frac{20\alpha^{3}l^{6}}{27\tau^{9}}+\cdots\label{eq: Solutions to holonomy conditions}
\end{equation}
 These solutions satisfy the ``integrability conditions'' (as can be checked order by order), 
\begin{eqnarray}
\frac{\partial L}{\partial\alpha}=\frac{\partial W}{\partial\tau}, \label{names}
\end{eqnarray}
 pointing out to the existence of a bulk (Euclidean) action, $I$
\begin{eqnarray}
\delta I\sim \delta \tau \  L+\delta \alpha \  W
\end{eqnarray}
 %so that the on-shell action,  \begin{eqnarray} I^{\mbox{on-shell}}\propto\tau L+\alpha W, \end{eqnarray}
with $L$ and $W$ being functions of $\tau,\alpha $. The basic reason why eqn (\ref{names}) arises is because we are demanding that there be an underlying partition function description for the system (the exponential of the action   being the semi-classical partition function). The integrability condition is the statement that the double derivatives of  the partition function (with respect to $\alpha$ and $\tau$) commute. A closely related discussion can be found in section (5.2) of \cite{Gutperle:2011kf}. The precise form of the action functional requires taking care of various subtleties, see \cite{David:2012iu}. We will make use of their results when we make comparisons with the AdS case.

For the (higher
spin) AdS case, the integrability conditions were understood \cite{Gutperle:2011kf}
to be integrability conditions of a $boundary$ CFT partition function,
$Z_{CFT}$ dual to the higher spin AdS bulk theory (vide AdS/CFT).
The on-shell bulk action, $I^{\mbox{on-shell}}$ is the saddle-point
contribution to $Z_{CFT}$, corresponding to the classical higher
spin Black hole configuration. Similarly it will be shown in Sec.
\ref{sec: dS TD}, the integrability conditions Eq. (\ref{eq:Holonomy constraints},\ref{eq: Solutions to holonomy conditions})
for the case of (higher spin) de Sitter connections apply to that
a putative dual $Euclidean$ CFT partition function, $Z_{CFT*}$.
It will also be shown that two partition functions ($Z_{CFT}$, $Z_{CFT*}$)
are related by a suitable ``Wick-rotation'' of the Cherns-Simons
level number (cosmological constant) and gauge theory charges (mass,
spin, spin-3 charges).

Finally we apply the radial gauge transformation, (\ref{eq: Radial gauge transformation})
to obtain full radial dependence, 
\begin{eqnarray}
A' & =iT_{0}d\rho+\left[\left(e^{\rho}-e^{-\rho}\frac{L}{2l}\right)T_{1}+i\left(e^{\rho}+e^{-\rho}\frac{L}{2l}\right)T_{2}+e^{-2\rho}\;\frac{W}{8l}W_{-2}\right]dw\nonumber \\
 & +\mu\left[e^{2\rho}\; W_{2}-\frac{L}{2l}W_{0}+e^{-2\rho}\;\frac{L^{2}}{16l^{2}}\; W_{-2}+e^{-\rho}\frac{W}{l}(T_{1}-iT_{2})\right]d\bar{w},\nonumber \\
\bar{A}' & =-iT_{0}d\rho+\left[\left(e^{\rho}-e^{-\rho}\frac{\bar{L}}{2l}\right)T_{1}-i\left(e^{\rho}+e^{-\rho}\frac{\bar{L}}{2l}\right)T_{2}+e^{-2\rho}\;\frac{\bar{W}}{8l}W_{-2}\right]d\bar{w}\nonumber \\
 & +\bar{\mu}\left[e^{2\rho}\; W_{2}-\frac{L}{2l}W_{0}+e^{-2\rho}\;\frac{L^{2}}{16l^{2}}\; W_{-2}+e^{-\rho}\frac{\bar{W}}{l}(T_{1}+iT_{2})\right]dw.
\end{eqnarray}
 \\
 The corresponding metric expression is 
\begin{eqnarray}
ds^{2} & = & -l^{2}d\rho^{2}+\left(\frac{lL}{2}+\frac{lW\bar{\mu}}{2}-\frac{1}{2}e^{-2\rho}L\overline{W}\bar{\mu}-\frac{\bar{L}^{2}\bar{\mu}^{2}}{3}\right)dw^{2}+\left(\frac{l\bar{L}}{2}+\frac{l\bar{W}\mu}{2}-\frac{1}{2}e^{-2\rho}\bar{L}W\mu-\frac{L^{2}\mu^{2}}{3}\right)d\bar{w}^{2}\nonumber \\
 &  & \left(\frac{1}{2}e^{2\rho}l^{2}-\frac{3lW\mu}{4}-\frac{3l\overline{W}\bar{\mu}}{4}+\frac{L^{2}\mu\bar{\mu}}{8}+\frac{L\bar{L}\mu\bar{\mu}}{12}+\frac{\bar{L}^{2}\mu\bar{\mu}}{8}+\frac{1}{8}e^{-2\rho}\left(L\bar{L}+4W\overline{W}\mu\bar{\mu}\right)\right)dwd\bar{w.}\label{eq: hs KdS_3 metric}
\end{eqnarray}
 \\
 while the expressions for the non-vanishing spin-3 field components
are, 
\begin{eqnarray}
\psi_{\rho\rho w} & = & \frac{1}{18}il^{2}\bar{L}\bar{\mu},\qquad\qquad\psi_{\rho\rho\bar{w}}=-\frac{1}{18}il^{2}L\mu,\nonumber \\
\psi_{www} & = & -\frac{1}{8}il^{2}W+\frac{1}{16}ilL^{2}\bar{\mu}+\frac{1}{24}ilL\bar{L}\bar{\mu}+\frac{1}{16}il\bar{L}^{2}\bar{\mu}-\frac{1}{12}il\bar{L}W\bar{\mu}^{2}+\frac{1}{27}i\bar{L}^{3}\bar{\mu}^{3}\nonumber \\
 &  & +\frac{1}{4}ie^{-2\rho}\left(lW\overline{W}\bar{\mu}-\frac{1}{6}L\bar{L}\overline{W}\bar{\mu}-\frac{1}{2}\bar{L}^{2}\overline{W}\bar{\mu}^{2}\right)-\frac{1}{8}ie^{-4\rho}W\overline{W}^{2}\text{\ensuremath{\bar{\mu}}}^{2}+\frac{ie^{-4\rho}\bar{L}^{2}\overline{W}^{2}\bar{\mu}^{3}}{16l}.\nonumber \\
\psi_{\bar{w}\bar{w}\bar{w}} & = & -\frac{1}{4}ie^{4\rho}l^{3}\mu+\frac{1}{2}ie^{2\rho}l^{2}W\mu^{2}-\frac{1}{24}ilL\bar{L}\mu+\frac{1}{12}ilL\overline{W}\mu^{2}-\frac{1}{27}iL^{3}\mu^{3}-\frac{1}{4}ilW^{2}\mu\nonumber \\
 &  & \qquad\qquad\qquad\qquad\qquad\qquad\qquad+\frac{1}{24}ie^{-2\rho}L\bar{L}W\mu^{2}+\frac{1}{32}ie^{-4\rho}\left(\bar{L}^{2}\overline{W}-\frac{L^{2}\bar{L}^{2}\mu}{2l}\right),\nonumber \\
\psi_{ww\bar{w}} & = & \frac{1}{12}ie^{2\rho}\left(l^{2}L\bar{\mu}+\frac{1}{3}il^{2}\bar{L}\bar{\mu}\right)\nonumber \\
 &  & +\frac{1}{24}il^{2}\overline{W}-\frac{1}{18}ilL^{2}\mu-\frac{1}{18}ilLW\mu\bar{\mu}-\frac{1}{72}iL^{2}\bar{L}\mu\bar{\mu}^{2}-\frac{1}{108}iL\bar{L}^{2}\mu\bar{\mu}^{2}-\frac{1}{72}i\bar{L}^{3}\mu\bar{\mu}^{2}\nonumber \\
 &  & \qquad+\frac{ie^{-2\rho}}{12}\left(\frac{1}{12}L\bar{L}^{2}\bar{\mu}+\frac{1}{4}\bar{L}^{3}\bar{\mu}-l\overline{W}^{2}\bar{\mu}-\frac{1}{2}l\bar{L}W+\frac{2}{3}L^{2}\overline{W}\mu\bar{\mu}+\frac{1}{36}\bar{L}W\overline{W}\mu\bar{\mu}^{2}\right)\nonumber \\
 &  & \qquad\qquad\qquad\qquad\qquad\qquad\qquad-\frac{ie^{-4\rho}}{24}\left(\frac{\bar{L}^{3}\overline{W}\bar{\mu}^{2}}{2l}-\bar{L}W\overline{W}\bar{\mu}-\overline{W}^{3}\bar{\mu}^{2}+\frac{L^{2}\overline{W}^{2}\mu\bar{\mu}^{2}}{2l}\right),\nonumber \\
\psi_{ww\bar{w}} & = & \frac{1}{12}ie^{4\rho}l^{3}\bar{\mu}-ie^{2\rho}l^{2}\left(\frac{1}{9}L\mu+\frac{1}{6}W\mu\bar{\mu}\right)\nonumber \\
 &  & +\frac{1}{12}i\left(lLW\mu^{2}+\frac{1}{6}l\bar{L}^{2}\bar{\mu}-\frac{1}{3}l\bar{L}\overline{W}\mu\bar{\mu}+\frac{1}{6}L^{3}\mu^{2}\bar{\mu}+\frac{1}{9}L^{2}\bar{L}\mu^{2}\bar{\mu}+\frac{1}{6}L\bar{L}^{2}\mu^{2}\bar{\mu}+lW^{2}\mu^{2}\bar{\mu}\right)\nonumber \\
 &  & \qquad\qquad\qquad\qquad\qquad\qquad\qquad+\frac{1}{12}ie^{-2\rho}\left(l\bar{L}\overline{W}-\frac{1}{3}L^{2}\bar{L}\mu-\frac{1}{6}\bar{L}^{2}W\mu\bar{\mu}-\frac{1}{3}LW\overline{W}\mu^{2}\bar{\mu}\right)\nonumber \\
 &  & \qquad\qquad\qquad\qquad\qquad\qquad\qquad\qquad-\frac{1}{24}ie^{-4\rho}\left(\bar{L}^{2}W-\frac{\bar{L}^{4}\bar{\mu}}{8l}+\bar{L}\overline{W}^{2}\bar{\mu}-\frac{L^{2}\bar{L}\overline{W}\mu\bar{\mu}}{2l}\right).\label{eq: Spin-3 field for hs KdS_3}
\end{eqnarray}

\subsection{Schwarzschild Gauge}

The Fefferman-Graham (FG) gauge expressions %
\begin{comment}

\subsubsection{pure SL(2,C)}

For pure $SL(2,C)$ connection, the horizon is at 
\begin{equation}
\rho_{+}=\frac{1}{2}\ln\frac{L}{2l}.\label{sl2hr}
\end{equation}
 Here we will specialize to the non-rotating solution 
\begin{equation}
L=\bar{L}
\end{equation}
 We define the Schwarzschild radial coordinate, $r$ by 
\begin{equation}
\rho=\ln\left[\frac{\left(r+\sqrt{r^{2}-r_{+}^{2}}\right)}{2l}\right],\label{red1}
\end{equation}
 where 
\begin{equation}
r_{+}^{2}=2lL.\label{r'}
\end{equation}
 It is convenient (for the purposes of making a double zero manifest
at the horizon) to go to $proper$ $radial$ coordinate, $x$ 
\begin{equation}
r^{2}=r_{+}^{2}\cosh^{2}x,
\end{equation}
 and write the metric as 
\begin{eqnarray}
ds^{2}=-l^{2}dx^{2}+\frac{2L\sinh^{2}x}{l}dt^{2}+2lL\cosh^{2}xd\phi^{2}
\end{eqnarray}
 
\end{comment}
only cover a part of the spacetime outside the horizon. In this section,
we describe solution of the gauge connection in Schwarzschild gauge.
For simplicity, we will consider the purely non-rotating case from
now on, 
\begin{eqnarray}
L=\bar{L},\qquad\bar{W}=-W\qquad\text{and}\qquad\bar{\mu}=-\mu.
\end{eqnarray}
 The metric in FG gauge is then, 
\begin{eqnarray}
g_{\rho\rho} & = & -l^{2},\nonumber \\
g_{tt} & = & \left(e^{\rho}-\frac{L+2W\mu}{2l}e^{-\rho}\right)^{2},\nonumber \\
g_{\phi\phi} & = & l^{2}\left(e^{\rho}+\frac{L-2W\mu}{2l}e^{-\rho}\right)^{2}+\frac{4L^{2}|\mu|^{2}}{3}-2l\, W\mu.
\end{eqnarray}
 We observe that there is a horizon i.e. $g_{tt}$ vanishes, at 
\begin{equation}
\rho_{+}=\frac{1}{2}\ln\left[\frac{L+2W\mu}{2l}\right].\label{sl3hr}
\end{equation}
 Now, we can introduce the Schwarzschild radial coordinate $r$ (motivated
from the definition of the Schwarzschild radial coordinate for the
pure $SL(2)$ case), 
\begin{eqnarray}
\rho=\ln\left[\frac{r+\sqrt{r^{2}-r_{+}^{2}}}{2l}\right],
\end{eqnarray}
 where $r_{+}^{2}$ is 
\begin{equation}
r_{+}^{2}=2l(L+2W\mu).
\end{equation}
 In the limit $\mu=0$, the above equation reduces to the pure $SL(2)$
case, (\ref{eq: FG radius in terms of Sch. radius}). In the Schwarzschild
like gauge the metric is given by, 
\begin{eqnarray}
ds^{2} & = & -\frac{l^{2}}{r^{2}-r_{+}^{2}}dr^{2}+\frac{2(L+2W\mu)}{l\, r_{+}^{2}}\left(r^{2}-r_{+}^{2}\right)dt^{2}+\nonumber \\
 &  & \left[\left(\frac{r\, L}{L+2W\mu}+\frac{2l\mu W\sqrt{r^{2}-r_{+}^{2}}}{L+2W\mu}\right)^{2}+\frac{4L^{2}|\mu|^{2}}{3}-2l\, W\mu\right]d\phi^{2}\label{eq:Sch. gauge higher spin}
\end{eqnarray}
 We also note that $g_{\phi\phi}>0$ as it is a sum of manifestly
positive quantities ($W$ and $\mu$ are imaginary quantities with
same sign vide (\ref{eq: Solutions to holonomy conditions})) and
there are no closed timelike curves in the $\phi$ direction.

\subsection{Higher Spin Quotient Cosmologies}

Now that we have the metric expressions for the higher spin versions
of the Kerr de Sitter universe in Schwarzschild gauge (\ref{eq:Sch. gauge higher spin})
outside the cosmological horizon, one can now write down metric for
higher spin generalizations of the quotient cosmologies (\ref{eq: Quotient Cosmology metric})
by simply swapping $r$ and $t$. 
\begin{equation}
ds^{2}=-\frac{l^{2}}{t^{2}-r_{+}^{2}}dt^{2}+\frac{2(L+2W\mu)}{l\, r_{+}^{2}}\left(t^{2}-r_{+}^{2}\right)dr^{2}+\left[\left(\frac{L\, t+2\mu W\sqrt{t^{2}-r_{+}^{2}}}{L+2W\mu}\right)^{2}+\frac{4L^{2}|\mu|^{2}}{3}-2l\, W\mu\right]d\phi^{2}.\label{eq: hs quotient cosmology}
\end{equation}
 Just as in the case for the $SL(2)$ quotient cosmology, $r$ is
now compactified into a circle and this metric cannot be continued
inside the horizon, $r_{+}$. As a result this it contains
big bang/big crunch like singularities at $t=\pm r_{+}$ when the
$r$-circle degenerates to a point, exactly like its $SL(2)$ cousin.
It will be interesting to consider the resolution of these singularities
along the lines of \cite{Krishnan:2013cra}, but we will not pursue
it here.

\section{Thermodynamics of asymptotically de Sitter connections\label{sec: dS TD}}

The aim of this section to derive a consistent thermodynamics for
asymptotically dS$_{3}$ spin-2 connections in the Chern-Simons language.
 (See \cite{Perez:2013xi} for an explicit expression for the entropy in metric-like variables.)
In a metric (second order) formalism of gravity, more precisely spin-2
gravity, thermodynamics of spacetimes containing horizons of any kind
is provided by the Gibbons-Hawking generalization \cite{Gibbons:1977mu}
of the black hole thermodynamics of Bardeen, Carter and Hawking \cite{Bardeen:1973gs}. 
%to include cosmological horizons. 
However, as we shall see, in the Chern-Simons or first order set-up,
a consistent thermodynamics is obtained extremely efficiently by first
mapping de Sitter solutions to Euclidean AdS (EAdS) solutions and
then demanding integrability conditions on free energy (equivalently
partition function) of a putative Euclidean CFT located on the future
infinity of the asymptotic de Sitter (same as conformal boundary of
the analytically continued EAdS solution). Maldacena \cite{Maldacena:2002vr}
notes that the conformal patch of dS %
\footnote{ie., the upper quadrant in the dS Penrose diagram containing the infinite
future at $\eta=0$ and bounded by the horizon at $\eta=1$. No light
rays from the infinite past can reach this region.%
}, 
\begin{eqnarray}
ds^{2}=\frac{-d\eta^{2}+d{\bf x}_{d}^{2}}{\eta^{2}/l^{2}}
\end{eqnarray}
 goes over to the Poincare patch of the EAdS, under, $l^{2}\rightarrow-l^{2}$
and, $\eta^{2}\rightarrow-z^{2}$, 
\begin{eqnarray}
ds^{2}=\frac{dz^{2}+d{\bf x}_{d}^{2}}{z^{2}/l^{2}},
\end{eqnarray}
 and then he proposes that for any asymptotic (in time) de Sitter
space, 
\begin{eqnarray}
\Psi_{Hartle-Hawking}=Z_{CFT^{*}},
\end{eqnarray}
since for EAdS one has the celebrated AdS-CFT conjecture $Z_{ESUGRA}=Z_{CFT}.$
Under the identifications, the Euclidean path-integral in AdS becomes
the Hartle-Hawking wave function of dS. %(however note due to the changein the sign of the cosmological constant the CFT dual for dS must have an $imaginary$ central charge, hence the asterisk in notation, $CFT^{*}$)
Next we construct a similar map between the $exterior$ regions of
Kerr deSitter and and Euclidean BTZ black hole and then generalize
to the higher spin case where the bulk action would be a first order
action instead of second order (metric) action.

\subsection{``Wick-rotation'' from Kerr de Sitter to $EBTZ$}

We simply write down these identifications for the FG gauge, 
\begin{eqnarray}
\rho_{dS} & \rightarrow & \rho_{EAdS}+i\:\frac{\pi}{2},\nonumber \\
t_{ds} & \rightarrow & i\; t_{EAdS}\\
l_{dS} & \rightarrow & i\: l_{EAdS},\nonumber \\
L_{dS},\bar{L}_{dS} & \rightarrow & -iL_{EAdS},-i\bar{L}_{EAdS}\nonumber \\
M_{dS},J_{dS} & \rightarrow & -M_{AdS},\;-J_{AdS}.\label{eq: dS to AdS identifications}
\end{eqnarray}
 Under these identifications, the KdS$_{3}$ metric Eq. (\ref{eq:Kerr-dS_3 outside})
goes over to 
\begin{equation}
ds^{2}\rightarrow d\tilde{s}^{2}=l^{2}d\rho^{2}+\frac{l}{2}\left(Ldw^{2}+\bar{L}d\bar{w}^{2}\right)+\left(l^{2}e^{2\rho}+\frac{L\bar{L}}{4}e^{-2\rho}\right)dwd\bar{w}\label{eq:Wick-rotated dS in FG}
\end{equation}
 but now with, 
\begin{eqnarray}
L=\frac{Ml+J}{2},\bar{L}=\frac{Ml-J}{2}.
\end{eqnarray}
 This is an evidently an Euclidean metric. To determine whether this
is the Euclidean $BTZ$ metric ($EBTZ$), we write the $EBTZ$ metric
expressions directly from Euclideanizing the Lorentzian $BTZ$, 
\begin{equation}
ds^{2}=\frac{l}{2}\left(L^{+}\: dw^{+2}+L^{-}\: dw^{-2}\right)+\left(l^{2}e^{2\rho}+\frac{L^{+}L^{-}}{4}e^{-2\rho}\right)dw^{+}\: dw^{-}+l^{2}d\rho^{2},w^{\pm}=\phi\pm\frac{t}{l},\label{eq: BTZ metric in FG}
\end{equation}
 where, the ``zero modes'' $L^{+},L^{-}$ are defined in terms of
the mass and the spin by, 
\begin{equation}
L^{+}=\frac{Ml+J}{2},L^{-}=\frac{Ml-J}{2}.\label{eq: BTZ zero modes}
\end{equation}
 Upon a replacing $t\rightarrow it_{E}$, we obtain the $EBTZ$ metric,
\begin{equation}
ds^{2}=\frac{l}{2}\left(L^{+}\: dw^{2}+L^{-}\: d\bar{w}^{2}\right)+\left(l^{2}e^{2\rho}+\frac{L^{+}L^{-}}{4}e^{-2\rho}\right)dw\: d\bar{w}+l^{2}d\rho^{2},w=\phi+\frac{it_{E}}{l},\bar{w}=\phi-\frac{it_{E}}{l},\label{eq:EBTZ in FG}
\end{equation}
 Clearly, this is identical to the wick-rotated $KdS_{3}$ metric
Eq. (\ref{eq:Wick-rotated dS in FG}).

\subsubsection{Schwarzschild Gauge Wick-rotation}

In Schwarzschild-like coordinates, the Kerr-dS$_{3}$ metric \ref{eq: Kerr-dS_3 in Sch},
on using the identifications, \ref{eq: dS to AdS identifications},
becomes, 
\begin{equation}
d\tilde{s}^{2}=N^{2}dt^{2}+N^{-2}dr^{2}+r^{2}\left(i\, N^{\phi}dt+d\phi\right)^{2},N^{2}=-M+\frac{r^{2}}{l^{2}}+\frac{J^{2}}{4r^{2}},N^{\phi}=\frac{J}{2r^{2}}.\label{eq:wick-rotated dS}
\end{equation}
 The Lorentzian exterior $BTZ$ metric Eq. (\ref{eq: BTZ metric in FG})
reads, 
\begin{equation}
ds^{2}=-N^{2}dt^{2}+N^{-2}dr^{2}+r^{2}\left(N^{\phi}dt+d\phi\right)^{2},N^{2}=-M+\frac{r^{2}}{l^{2}}+\frac{J^{2}}{4r^{2}},N^{\phi}=\frac{J}{2r^{2}}.\label{eq:BTZ in Sch}
\end{equation}
 which upon Euclideanizing i.e. $t\rightarrow it_{E},$ 
\begin{equation}
ds^{2}=N_{E}^{2}dt_{E}^{2}+dr^{2}/N_{E}^{2}+r^{2}\left(i\: N_{E}^{\phi}dt_{E}+d\phi\right)^{2},N_{E}^{2}=-M+\frac{r^{2}}{l^{2}}+\frac{J^{2}}{4r^{2}},N_{E}^{\phi}=\frac{J}{2r^{2}}.\label{eq:EBTZ in Sch}
\end{equation}
 One can write a metric expression in terms of outer and inner horizons
for the $BTZ$ along the lines of \ref{eq:Kerr-dS_3 outside}, 
\begin{equation}
ds^{2}=-\frac{\left(r^{2}-r_{+}^{2}\right)\left(r^{2}-r_{-}^{2}\right)}{r^{2}l^{2}}dt^{2}+\frac{r^{2}l^{2}}{\left(r^{2}-r_{+}^{2}\right)\left(r^{2}-r_{-}^{2}\right)}dr^{2}+r^{2}\left(d\phi+\frac{r_{+}r_{-}}{r^{2}}\frac{dt}{l}\right)^{2},r>r_{+}\label{eq: BTZ outside in Sch}
\end{equation}
 with, 
\begin{eqnarray}
r_{\pm}^{2}=\frac{Ml^{2}}{2}\left(1\pm\sqrt{1-\left(\frac{J}{Ml}\right)^{2}}\right).
\end{eqnarray}
 So the identifications are, 
\begin{equation}
r_{+}\rightarrow r_{+},\left(r_{-}\right)_{KdS_{3}}\rightarrow-i\,\left(r_{-}\right)_{BTZ}.\label{eq:r_plus and r_minus identifications}
\end{equation}
 For $KdS_{3}$ note that in terms of $L,\bar{L}$, 
\begin{eqnarray}
r_{+}=\frac{\sqrt{Ll}+\sqrt{\bar{L}l}}{\sqrt{2}},r_{-}=\frac{\sqrt{Ll}+\sqrt{\bar{L}l}}{\sqrt{2}}
\end{eqnarray}
 
\begin{eqnarray}
r_{+}^{2}+r_{-}^{2}=2\sqrt{L\bar{L}l^{2}}.
\end{eqnarray}
 So, the temperature inverse of $KdS_{3}$ in terms of $L,\bar{L}$,
\begin{equation}
\frac{\beta}{2\pi}=\frac{l^{2}}{2}\left(\frac{1}{\sqrt{2Ll}}+\frac{1}{\sqrt{2\bar{L}l}}\right).\label{eq: Inverse temperature as a function of L and Lbar}
\end{equation}
 For non-rotating $KdS_{3}$, $\tau=\beta/2\pi$ and we have 
\begin{eqnarray}
\frac{L}{l}=\frac{l^{2}}{2\tau^{2}}.
\end{eqnarray}
 Again, this metric (\ref{eq:EBTZ in Sch}) is exactly that of the
Wick-rotated dS metric in Schwarzschild coordinates, Eq. (\ref{eq:wick-rotated dS}).

\subsection{dS-AdS ``wick rotation'' at work: Equivalence of thermodynamics
in the metric formulation}

In order to further solidify our heuristic identifications, we show
that under these identifications the Gibbons-Hawking thermodynamics
\cite{Gibbons:1977mu}, including the temperature and entropy of the
Kerr-dS$_{3}$ solution, maps onto to those the ``wick-rotated''
EBTZ solutions. %as expected from \cite{Bardeen:1973gs}. 

\begin{enumerate}
\item The entropies for either geometry are the same since entropy of either
cosmological or Black-hole horizons in the Gibbons-Hawking framework
is given by 
\begin{equation}
S=\frac{1}{4G}(\mbox{Horizon Area})=2\left(2\pi r_{+}\right)=4\pi r_{+}.\label{eq: KdS_3 or BTZ entropy}
\end{equation}
 This is borne out by our heuristic identifications, since $r_{+}\rightarrow r_{+}$
. 
\item The temperature of KdS$_{3}$ is given by Gibbons-Hawking thermodynamics
by the conical singularity trick, 
\begin{equation}
T_{KdS_{3}}=\frac{r_{+}^{2}+r_{-}^{2}}{2\pi l^{2}r_{+}}.\label{eq: Hawking temperature of Kerr-dS_3}
\end{equation}
 Using the identification Eq. (\ref{eq:r_plus and r_minus identifications})
and the additional identification $T_{ds}\rightarrow-T_{AdS}$%
\footnote{This temperature sign flip is a direct result of the flip in the sign
of mass parameter or ``internal energy'' $M$ in identification
\ref{eq: dS to AdS identifications}. The conjugacy relation 
\begin{eqnarray}
T^{-1}=\frac{\partial S}{\partial M},
\end{eqnarray}
 directs you that one needs to perform, 
\begin{eqnarray}
T_{dS}\rightarrow-T_{AdS}
\end{eqnarray}
 in consonance with 
\begin{eqnarray}
M_{dS}\rightarrow-M_{AdS}.
\end{eqnarray}
} this temperature continues to the Hawking temperature of the corresponding
BTZ black hole! 
\begin{equation}
T_{BTZ}=\frac{r_{+}^{2}-r_{-}^{2}}{2\pi l^{2}r_{+}}.\label{eq: Hawking Temperature of BTZ}
\end{equation}

\item The chemical potential conjugate to angular momentum is, 
\begin{equation}
\Omega_{KdS_{3}}=-T\frac{\partial S}{\partial J}=-\frac{r_{-}}{r_{+}l}\label{eq: KdS_3 angular chemical potential}
\end{equation}
 Again under the identifications, we obtain the expected behavior
$\Omega_{dS}\rightarrow\Omega_{AdS}$ since $J_{dS}\rightarrow-J_{AdS}$%
\footnote{Since, going over from dS to AdS implies the replacements $S\rightarrow S$,
$J\rightarrow-J$ and $\beta\rightarrow-\beta$, we must have $\Omega\rightarrow\Omega$
in order to reproduce the correct thermodynamic relation, 
\begin{eqnarray}
\frac{\partial S}{\partial J}=-\Omega\beta
\end{eqnarray}
}. We note that, $\Omega_{BTZ}=\frac{r_{-}}{r_{+}l}.$ Parenthetically,
we note that when we move to Euclidean BTZ, we need to define, $J_{EAdS}=-iJ_{AdS}$
and consequently the new conjugate $\Omega_{EAdS}=i\Omega_{AdS}$,
so that respective identifications are, $J_{dS}\rightarrow-iJ_{EAdS}$
and $\Omega_{KdS_{3}}\rightarrow-i\Omega_{EBTZ}$. 
\end{enumerate}
Since under the identifications, one can successfully map any dS thermodynamic
quantities like entropy, internal energy, angular charges and their
respective conjugates to AdS quantities, the laws of thermodynamics
will continue as well. %\begin{eqnarray} \left.TdS=dE-\Omega dJ-\mu dQ\right|_{dS}\rightarrow\left.TdS=dE-\Omega dJ-\mu dQ\right|_{AdS} \end{eqnarray} provided we identify the charges $Q_{dS}\stackrel{?}{\rightarrow}Q_{AdS}$ and their conjugate potentials $\mu_{dS}\stackrel{?}{\rightarrow}\mu_{AdS}$.
When higher spin charges are added, we will demand a similar statement
to hold with higher spin charges and chemical potential added to the
thermodynamical relations.

\subsection{Thermodynamics in the Chern-Simons formulation}

So far everything we discussed was in the $SL(2,C)$ sector of the
theory with just metric or spin-2 fields turned on, but we extend
this analogy to the full $SL(3,C)$ sector i.e. when both metric and
spin-3 field are present. In that case though we do not know the generalization
of the Gibbons-Hawking thermodynamics \cite{Bardeen:1973gs}. However,
taking dS/CFT as a principle, we can propose that the thermodynamics
of a dS-connection is identical to that of a suitably continued Euclidean
AdS-connection i.e. a higher spin AdS black hole \cite{Gutperle:2011kf}.
The thermodynamics of $SL(3,C)$ valued Euclidean AdS$_{3}$-connections
for higher spin black-holes (connected to BTZ, i.e. the so called
``BTZ'' branch) has been shown to be dictated by the integrability
conditions of the free energy of a dual CFT \cite{Gutperle:2011kf}.
These conditions which can be cast in a gauge-invariant form by the
$holonomy$ conditions \cite{Gutperle:2011kf}. Under the correct
identifications of charges and potentials, the integrability or holonomy
conditions of a dS connection should continue to those of an AdS connection.
$\emph{\ensuremath{}}$Or turning this fact around, we expect the
charges we obtain functions of the potentials $\mu$ and $T$ on solving
the integrability conditions on the dS side, (\ref{eq: Solutions to holonomy conditions})
to reproduce the respective solution of AdS integrability conditions
i.e. AdS charges as a function of AdS potentials \cite{David:2012iu}
upon making the dS-to-AdS identifications. For AdS, the solution to
the holonomy conditions is,%
\begin{comment}
\begin{eqnarray}
\mathcal{L}=-\frac{k}{8\pi\tau^{2}}+\frac{5}{6}\frac{k}{\pi}\frac{\alpha^{2}}{\tau^{6}},\qquad\mathcal{W}=-\frac{k}{3\pi}\frac{\alpha}{\tau^{5}}+O(\alpha^{3}).
\end{eqnarray}
 
\end{comment}
\begin{equation}
\frac{L^{+}}{l}=\frac{l^{2}}{2\tau^{2}}+\frac{10}{3}\frac{\alpha^{2}l^{4}}{\tau^{6}}+\dots,\qquad\frac{W^{+}}{l}=-\frac{4}{3}\frac{\alpha l^{4}}{\tau^{5}}+\dots.\label{eq: Equation of State for AdS}
\end{equation}
{} {} {} We have the identifications for the spin-3 charges when going
from $dS$ to $AdS$, 
\begin{eqnarray}
\alpha_{AdS}=\frac{\alpha_{dS}}{2},W_{AdS}^{+}=-2iW_{dS},
\end{eqnarray}
 or, 
\begin{equation}
\mu_{AdS}=-\frac{\mu_{dS}}{2},W_{AdS}^{+}=-2iW_{dS}.\label{eq:Identifications for the spin-3 charge}
\end{equation}

\subsubsection{The Action and Free energy in the CS theory}

To compute, $Z_{CFT}$ from the bulk gauge theory, we make use of
the saddle point approximation, 
\begin{eqnarray}
Z_{CFT}=Z_{(E)SUGRA}=e^{I_{E}^{On-shell}}
\end{eqnarray}
 where, $I_{E}$ is the Euclidean bulk action, defined in terms of
the original action, $I$ by 
\begin{eqnarray}
I_{E}[F({\bf x},t)]=iI[F({\bf {\bf x}},it_{E})].
\end{eqnarray}
 The Chern-Simons action without any supplementary boundary terms,
\begin{eqnarray}
I_{CS}=\frac{k}{4\pi\epsilon_{R}}\int\:\mbox{Tr}\left(AdA+\frac{2}{3}A^{3}\right),\epsilon_{R}=4,
\end{eqnarray}
 is the right action for the $SL(2)$ sector. On-shell this becomes
\cite{Banados:1998ta}, 
\begin{equation}
I_{CS}[A]=-\frac{k}{4\pi\epsilon_{R}}\int dt\, d\mbox{\ensuremath{\phi}}\; Tr\left(A_{t}A_{\phi}\right).\label{eq: On shell CS action}
\end{equation}
 For $SL(3)$ sector one needs to add new boundary terms as formulated
in \cite{Banados:2012ue,deBoer:2013gz}. But it is easy to see that
a similar map as we are presenting below will also hold for the boundary
terms, so in the following, we will illustrate it only for the bulk
terms. %the uncorrected action,
%$I_{CS}$ is satisfactory enough to bring out the isomorphism between the thermodynamics of the AdS black holes and KdS$_{3}$ spaces and their higher spin versions since all boundary terms are based on this and hence must require AdS-to-dS ``wick-rotatable'' terms. 
Using, (5.1) of \cite{Gutperle:2011kf} %
\begin{comment}
we have, 
\begin{eqnarray}
\mbox{Tr}\left(a_{t}a_{\phi}\right)=\frac{16\pi\mathcal{L}}{k}-\frac{512\pi^{2}\mu^{2}\mathcal{L}^{2}}{3k^{2}}
\end{eqnarray}
 
\begin{eqnarray}
\mbox{Tr}\left(\bar{a}_{t}\bar{a}_{\phi}\right)=-\frac{16\pi\bar{\mathcal{L}}}{k}+\frac{512\pi^{2}\bar{\mu}^{2}\bar{\mathcal{L}}^{2}}{3k^{2}}.
\end{eqnarray}
 So the on-shell euclidean action%
\footnote{Note that we need to put $k/16\pi$ instead of the $k/4\pi$ when
defining the CS action, 
\begin{eqnarray}
\frac{k}{16\pi}\int\left(AdA+\frac{2}{3}A^{3}\right)
\end{eqnarray}
 and our conventions for normalizations etc. are that of \cite{deBoer:2013gz}. %
} for the non-rotating case, $\mathcal{L}=\bar{\mathcal{L}},\mu=-\bar{\mu}$,
\begin{eqnarray*}
I_{\mbox{on-shell}} & = & i\left[-\frac{k}{4\pi\epsilon_{R}}\int dtd\phi\left(\frac{16\pi\mathcal{L}}{k}-\frac{512}{3}\frac{\pi^{2}\mu^{2}\mathcal{L}^{2}}{k^{2}}\right)\times2\right],\epsilon_{R}=4\\
 & = & i\left[-\frac{k}{16\pi}\int_{0}^{-\beta}i\, dt_{E}\int_{0}^{2\pi}d\phi\left(\frac{16\pi\mathcal{L}}{k}-\frac{512}{3}\frac{\pi^{2}\mu^{2}\mathcal{L}^{2}}{k^{2}}\right)\times2\right],t=it_{E}\\
 & = & -4\pi\beta\mathcal{L}+\frac{128\pi^{2}\beta\mu^{2}\mathcal{L}^{2}}{3k}.
\end{eqnarray*}
 Now, substituting the notation changes, 
\begin{eqnarray}
\frac{2\pi}{k}\mathcal{L}=\frac{L}{2l},k=2l\implies\mathcal{L}=\frac{L}{2\pi},k=2l.
\end{eqnarray}
 we get, 
\end{comment}
\begin{equation}
I_{\mbox{on-shell}}^{\mbox{EAdS}}=-\frac{2\beta L}{l}+\frac{16\beta\mu^{2}L^{2}}{3l^{2}}.\label{eq:On-shell action for AdS}
\end{equation}
{} {} %
\begin{comment}
Now, lets evaluate the on-shell action for the de Sitter side. We
need, 
\begin{eqnarray}
\mbox{Tr}\left(A_{t}A_{\phi}\right)=-\frac{4iL}{l^{2}}-\frac{8i\mu^{2}L^{2}}{3l^{3}},
\end{eqnarray}
 
\end{comment}
The higher spin de Sitter on-shell action turns out to be%
\footnote{To get this on-shell action, we perform integration over $t$-circle
$(0,i\beta)$ and over $\phi$-circle $(0,2\pi)$ 
\begin{eqnarray}
\tilde{I}_{\mbox{on-shell }}^{\mbox{dS}}=i\left(-\frac{k}{4\pi\epsilon_{R}}\right)\int_{0}^{-\beta}\left(i\, dt\right)\int_{0}^{2\pi}d\phi\left[-i\:\mbox{Tr}\left(A_{t}A_{\phi}-\bar{A}_{t}\bar{A}_{\phi}\right)\right],
\end{eqnarray}
 with 
\begin{eqnarray}
k=-2il.
\end{eqnarray}
}, 
\begin{equation}
\tilde{I}_{\mbox{on-shell}}^{\mbox{dS}}=-\left(\frac{2\beta L}{l}+\frac{4}{3}\frac{\beta\mu^{2}L^{2}}{l^{2}}\right).\label{eq: On-shell action}
\end{equation}
 Again, using the identifications, we see that the $dS$ on-shell
action reproduces the $EAdS$ on-shell action, (\ref{eq:On-shell action for AdS})
\begin{eqnarray}
\tilde{I}_{\mbox{On-shell}}^{\mbox{dS}}=I_{\mbox{On-shell}}^{\mbox{EAdS}}.
\end{eqnarray}
Thus we have demonstrated that the higher spin generalizations of
Kerr de Sitter universes %have a thermodynamics identical to that of
are related to (higher spin) AdS black holes just as they were in the pure gravity
(spin-2) case in the metric formulation. However this on-shell action is not yet equal to a $-\beta \Phi $, where $\Phi$ is the grand ``higher spin" canonical potential $ \Phi = E- TS - \mu W$. But it is possible to add boundary/supplementary terms and change the action $\tilde{I}_{\mbox{on-shell}}^{\mbox{dS}}$ to a new action $I_{\mbox{On-shell}}$ such that,
\begin{eqnarray}
-I_{\mbox{On-shell}}= \beta \Phi.
\end{eqnarray}
Such a procedure was conducted in the anti de Sitter case in \cite{David:2012iu}, see their section (2.2). For our de Sitter case, the necessary extra terms can be obtained from their expressions
by the AdS-dS identifications (51) and (77), in
exact analogy with our computation here for the bulk terms. The match
between our entropy and the higher spin AdS$_{3}$ black hole entropy
\cite{Kraus:2013esi} is a natural consequence, and we have explicitly
checked this. This concludes our discussion about the connection between
the thermodynamics of the Kerr-dS$_{3}$ solution and that of higher
spin black holes in AdS$_{3}$.

\section*{Acknowledgments}

We thank Justin David for clarifications on \cite{David:2012iu}.
CK thanks Oleg Evnin for discussions on higher spins and singularity
resolution, and Oleg Evnin and Auttakit Chatrabhuti for hospitality
at Chulalangkorn University, Bangkok, during part of this project.
The research of SR is supported by Department of Science and Technology
(DST), Govt. of India research grant under scheme DSTO/1100 (ACAQFT).

 \bibliographystyle{brownphys}
\bibliography{hscosmo}
 
\end{document}